\documentclass{pnastwo}
\usepackage[dvips]{graphicx}
\usepackage{amssymb,amsfonts,amsmath}
\usepackage{float}

\def\apj{{\it Astrophys.~J.\ }}
\def\apjl{{\it Astrophys.~J.~Lett.\ }}

\def\mnras{{\it Mon.~Not.~R.~Astron.~Soc.\ }}

\def\apjs{{\it Astrophys.~J.~Suppl.\ }}

\def\aap{{\it Astron. Astrophys.}\ }

\begin{document}

\title{Highlights in the Study of Exoplanet Atmospheres}  

\author{Adam S. Burrows\affil{1}{Department of Astrophysical Sciences, Princeton University, Princeton, NJ 08544}}

\maketitle

\begin{article}

\begin{abstract}

{\bf Exoplanets are now being discovered in profusion.  However, to understand
their character requires spectral models and
data. These elements of remote sensing can yield temperatures,
compositions, and even weather patterns, but only if significant improvements
in both the parameter retrieval process and measurements are achieved.
Despite heroic efforts to garner constraining data on
exoplanet atmospheres and dynamics, reliable interpretation has oftimes lagged
ambition. I summarize the most productive, and at times novel, methods
employed to probe exoplanet atmospheres, highlight some of the most interesting
results obtained, and suggest various broad theoretical topics in which further
work could pay significant dividends.}

\end{abstract}

\keywords{exoplanets | atmospheres | planetary science | spectroscopy | characterization}

\section{Introduction}
\label{intro}

The modern era of exoplanet research started in 1995 with 
the discovery of the planet 51 Peg b \cite{mayor} due to the 
detection of the periodic radial-velocity (RV) Doppler wobble in its star (51 Peg) 
induced by the planet's nearly circular orbit. With these data, 
and knowledge of the star, one could derive orbital period ($P$) and semi-major axis ($a$), 
and constrain the planet's mass.  However, the inclination of the planet's orbit
was unknown and, therefore, only a lower limit to its mass could be determined.
With a lower limit of 0.47 M$_{\rm J}$ (where M$_{\rm J}$ is the mass of Jupiter),
and given its proximity to its primary ($a$ is $\sim$0.052 A.U.; one hundred times
closer to its star than Jupiter is to the Sun), the induced Doppler wobble is 
optimal for detection by the RV technique.  The question was how such a 
``hot Jupiter" could exist and survive.  While its survival is now 
understood (see section below on ``Winds from Planets"), the reason for its close 
orbital position is still a subject of vigorous debate.  Nevertheless, such 
close-in giants are selected for using the RV technique and soon scores, 
then hundreds, of such gas giants were discovered in this manner. 

However, aside from a limit on planet mass, and the inference that proximity to its star 
leads to a hot ($\sim$1000-2000 Kelvin (K)) irradiated atmosphere, no useful physical information on such planets
was available with which to study planet structure, their atmospheres, or composition.
A breakthrough along the path to characterization and the establishment of a mature
field of exoplanet science occurred with the discovery of giant planets, still 
close-in, that {\it transit} the disk of their parent star.  The chance of a transit is larger
if the planet is close and HD 209458b at $a$$\sim$0.05 A.U. was the first found \cite{first}.  
Optical measurements yielded a radius (R$_{p}$) for HD 209458b of $\sim$1.36 R$_{\rm J}$, where R$_{\rm J}$ is the radius 
of Jupiter.  Jupiter is roughly ten times, and Neptune is roughly four times, the radius of Earth 
(R$_{\rm E}$). Since then, hundreds of transiting giants have been discovered using ground-based facilities.
The magnitude of the attendant diminution of a star's light during such a primary transit 
(eclipse) by a planet is the ratio of their areas ($\frac{R_p^2}{R_*^2}$, where R$_{p}$ and R$_{*}$ are
the planet's and star's radius, respectively), so with knowledge of the star's radius
the planet's radius can be determined.  Along with RV data, since the orbital inclination of a planet
in transit is known, one then has a radius$-$mass pair with which to do some science.   
The transit depth for a giant passing in front of a solar-like star is $\sim$1\%, and such a large
magnitude can easily be measured with small telescopes from the ground.  A smaller Earth-like
planet requires the ability to measure transit depths one hundred times more precisely.
Soon, many hundreds of gas giants were detected both in transit and via the RV method, the former 
requiring modest equipment and the latter requiring larger telescopes with state-of-the-art
spectrometers with which to measure the small stellar wobbles.   Both techniques
favor close-in giants, so for many years these objects dominated the beastiary of
known exoplanets. 

Better photometric precision near or below one part in $10^{4-5}$, achievable only 
from space, is necessary to detect the transits of Earth-like and Neptune-like exoplanets 
across Sun-like stars, and, with the advent of the {\it Kepler} \cite{borucki_2010} and 
CoRoT \cite{corot} satellites, astronomers have now discovered a few thousand exoplanet 
candidates. {\it Kepler} in particular revealed that most planets are smaller than 
$\sim$2.5 R$_{\rm E}$ (four times smaller than Jupiter), but fewer than $\sim$100 of the {\it Kepler} candidates 
are close enough to us to be measured with state-of-the-art RV techniques.  Without masses, 
structural and bulk compositional inferences are problematic.  Moreover, the 
majority of these finds are too distant for photometric or spectroscopic follow-up from the 
ground or space to provide thermal and compositional information.  

A handful of the {\it Kepler} and CoRoT exoplanets, and many of the transiting giants
and ``sub-Neptunes" discovered using ground-based techniques are
not very distant and have been followed up photometrically and spectroscopically 
using both ground-based and space-based assets to help constrain their atmospheric properties.
In this way, and with enough photons, some information on atmospheric compositions 
and temperatures has been revealed for $\sim$50 exoplanets, mostly giants.  However, even 
these data are often sparse and ambiguous, rendering most such hard-won results 
provisional \cite{burrows_pnas}. The nearby systems hosting larger transiting 
planets around smaller stars are the best targets for a program of remote sensing to be 
undertaken, but such systems are a small subset of the thousands of exoplanets 
currently in the catalogues.  

One method with which astronomers are performing such 
studies is to measure the transit radius as a function of wavelength 
\cite{brown_transit,seager_sasselov,fortney_2003}.  Since the opacity of molecules
and atoms in a planet's atmosphere is a function of wavelength, the apparent size
of the planet is a function of wavelength as well, in a manner characteristic
of atmospheric composition.  Such a ``radius spectrum" can reveal the atmosphere's
composition near the planet terminators, but the magnitude of the associated variation 
is down from the average transit depth by a factor of $\sim$$\frac{2H}{R_p}$, where $H$ is 
the atmospheric scale height (a function of average temperature and gravity).  This ratio 
can be $\sim$0.1 to 0.01, making it correspondingly more difficult to determine 
a transit radius spectrum.  Only space telescopes such as {\it Spitzer} \cite{werner} 
and the Hubble Space Telescope (HST), and the largest ground-based telescopes with advanced spectrometers, 
are up to the task, and even then the results can be difficult to interpret.

Another method probes the atmospheres of transiting exoplanets at secondary eclipse, 
when the star occults the planet $\sim$180$^{\circ}$ out of phase with the primary transit.
The abrupt difference between the summed spectrum of planet and star just before and during the eclipse
of the planet by the star is the planet's spectrum at full face. Secondary eclipse spectra
include reflected (mostly in the optical and near-ultraviolet) and thermally emitted (mostly
in the near- and mid-infrared) light, and models are necessary to distinguish (if possible) 
the two components. Note that separate images of the planet and star are not obtained
via this technique, and a planet must be transiting.  With few exceptions, when the planet does not 
transit the summed light of a planet and star varies too slowly and smoothly for such a variation 
to be easily distinguished from the systematic uncertainties of the instruments to reveal
the planet's emissions as a function of orbital phase. For the close-in transiting 
``hot Jupiters," the planet flux in the near-infrared is $\sim$10$^{-3}$ times the stellar flux, much 
higher than the ratio expected for the class of planet in a wide orbit that 
can be separated from its primary star by high-contrast imaging techniques. 
In cases when such ``high-contrast" direct imaging is feasible, the planet is farther away from
the star (hence, dim) and difficult to discern from under the stellar glare. 
However, hot, young giants can be self-luminous enough to be captured by current 
high-contrast imaging techniques and a handful of young giant planets have been
discovered and characterized by this technique. More are expected as the technology matures \cite{GPI,SPHERE,suzuki,spergel}.

The secondary eclipse and primary transit methods used to determine or constrain atmospheric 
compositions and temperatures (as well as other properties) generally involve low-resolution
spectra with large systematic and statistical errors. These methods are complementary in that 
transit spectra reliably reveal the presence of molecular and atomic features and are 
an indirect measure of temperature through the pressure scale height, while the 
flux levels of secondary eclipse spectra scale directly with temperature, but could in 
fact be featureless for an isothermal atmosphere. The theoretical spectra with which they
are compared to extract parameter values are imperfect as well, and this results in less 
trustworthy information than one would like.  Giant planets (and ``Neptunes") orbiting 
closely around nearby stars are the easiest targets, and are the stepping stones 
to the Earths.  Secondary and primary transit spectral measurements of Earth-like planets 
around Sun-like stars, as well as direct high-contrast imaging of such small planets, 
are not currently feasible.  However, measurements of exo-Earths around smaller
M dwarf stars might be, if suitable systems can be found.  Nevertheless, with a few score transit and 
secondary eclipse spectra, some planetary phase light curves, a few high-contrast campaigns
and measurements, and some narrow-band, but very high spectral resolution measurements
using large telescopes, the first generation of exoplanet atmosphere studies has begun. 

There are several helpful reviews of the theory of exoplanet atmospheres 
\cite{madhu_exo,burrows_orton,fletcher,tinetti,echo2,guillot_2010,burrows_rmp_2001,burrows_1997,deming_JWST}.
To these can be added informed discussions on the molecular spectroscopy and opacities 
central to model building \cite{HITRAN,tennyson,exomol,freedman,sharp_burrows}. 
Monographs on the relevant thermochemistry and abundances have been published over the years 
\cite{lodders_2003,lodders_fegley,lodders_fegley_companion,burrows_sharp,fegley_silicate_atmos}.
In this paper, I will not attempt to review the literature of detections and claims, nor will
I attempt to review the thermochemical, spectroscopic, or dynamical modeling efforts to date.
Rather, I will focus on those few results concerning exoplanet atmospheres that to my mind
stand out, that seem most robust, and that collectively serve to summarize what we have truly learned.
This will of necessity be a small subset of the published literature, and, if only 
for lack of space, some compelling results will no doubt be neglected. In addition, I will
touch on only the basics of the atmosphere theory applied to date, preferring to focus where
possible on the progress in theory necessary for the next generation of exoplanet atmosphere studies
to evolve productively.  I now embark upon a discussion of what I deem a few of the milestone 
observational papers in core topics. These might be considered to constitute the 
spine of progress in recent exoplanet atmosphere studies.  I accompany each with a 
short discussion of the associated theoretical challenges posed by the data.

\section{Transit Detection of Atoms and Molecules}
\label{transit}

The apparent transit radius of a planet with a gaseous atmosphere is 
that impact parameter of a ray of stellar light for which the optical depth
at that wavelength ($\lambda$) is of order unity. Note that at that level 
the corresponding radial optical depth, which if in absorption is 
relevant to emission spectra at secondary eclipse, will be much smaller.  
Since an atmosphere has a thickness (extent) and absorption and scattering 
cross sections are functions of photon wavelength that in product 
with the air column constitute optical depth, the measured transit 
radius is a function of wavelength.  Therefore, measurements of a 
planet's transit depths at many wavelengths of light reveal its atomic and 
molecular composition.  To good approximation \cite{lecavelier}:
\begin{equation}
\frac{dR_{\rm p}}{d\ln{\lambda}} \approx H\frac{d\ln{\sigma(\lambda)}}{d\ln{\lambda}} \, ,
\label{sigma_r}
\end{equation}
where $\sigma(\lambda)$ is the composition-weighted total cross-section and the scale height, $H$, is 
$kT/\mu g$, where $g$ is the planet's surface gravity, $\mu$ is the mean molecular weight, 
$T$ is an average atmospheric temperature, and $k$ is Boltzmann's constant. $H$ sets
the scale of the magnitude of potential fluctuations of R$_{\rm p}$ with $\lambda$ and $\sigma(\lambda)$
is determined mostly by the atomic and molecular species in the atmosphere.

Charbonneau et al. \cite{sodium} were the first to successfully employ this technique with
the $\sim$4$-$$\sigma$ measurement of atomic sodium (Na) in the atmosphere of HD 209458b.   
Along with HD 189733b, this nearby giant planet has been the most scutinized photometrically
and spectroscopically. Since then, Sing et al. \cite{sing_pot} have detected potassium (K)
in XO-2b and Pont et al. \cite{pont2} have detected both sodium and potassium in HD 189733b. 
These are all optical measurements at and around the Na D doublet ($\sim$0.589 $\mu$m) and the potassium
resonance doublet ($\sim$0.77 $\mu$m), and the measurements revealed the telltale 
differential transit depths in and out of the associated lines.

From experience with brown dwarfs, the presence of neutral alkali metals in the atmospheres
of irradiated exoplanets with similar atmospheric temperatures ($\sim$1000$-$1500 K) was expected
and their detection was gratifying.  Indeed, there is a qualitative correspondence between the 
atmospheres of close-in/irradiated or young giant planets (of order Jupiter's mass) 
and older brown dwarfs (with masses of tens of M$_{\rm J}$).  Alkalis persist to 
lower temperatures ($\sim$800-1000 K) to be revealed in close-in exoplanet transit and emission spectra and 
in older brown dwarf emission spectra because the silicon and aluminum with which they would 
otherwise combine to form feldspars are sequestered at higher temperatures and depths 
into more refractory species and rained out. Had the elements with which Na and K would have 
combined persisted in the atmosphere at altitude, these alkalis would have combined and 
their atomic form would not have been detected \cite{bur_mar}.  The more refractory silicates (and condensed iron) 
reside in giant exoplanets (and in Jupiter and Saturn), but at great depths.  In L dwarf 
brown dwarfs they are at the surface, reddening the emergent spectra significantly.

However, the strength in transiting giant exoplanets of the contrast in and out of these atomic alkali lines 
is generally less than expected \cite{fortney_2003}.  Subsolar elemental Na and K abundances, ionization
by stellar light, and hazes have been invoked to explain the diminished strength of their associated 
lines, but the haze hypothesis is gaining ground.  The definition of a haze can merge with that 
of a cloud, but generally hazes are clouds of small particulates at altitude that may be condensates
of trace species or products of photolysis by stellar UV light and polymerization.  They are generally
not condensates of common or abundant molecular species (such as water, ammonia, iron, or silicates,
none of which fit the bill here).  Though what this haze is is not at all clear,
hazes at altitude ($\le$0.01 bars) can provide a nearly featureless continuum opacity to light 
and easily mute atomic and molecular line strengths. Indeed, hazes are emerging as central and 
ubiquitous features in exoplanet atmospheres.  Annoyingly, not much mass is necessary to have an effect
on transit spectra, making quantitative interpretation all the more difficult.  The fact that
the red color of Jupiter itself is produced by a trace species (perhaps a haze) that as of
yet has not been identified is a sobering testament to the difficulties that lie ahead in
completely determining exoplanet atmospheric compositions.

The multi-frequency transit measurements of HD 189733b performed by Pont et al. \cite{pont2,pont} 
from the near-ultraviolet to the mid-infrared are the clearest and most dramatic
indications that some exoplanets have haze layers (Figure 1).  Curiously, no water or 
other molecular features are identified by Pont et al. \cite{pont2,pont} in transit. Aside 
from the aforementioned Na and K atomic features in the optical, the transit spectrum of 
HD 189733b is consistent with a featureless continuum.  Water features
in a H$_2$ atmosphere are very difficult to completely suppress, 
so this is strange. What is more, the transit radius increases below 
$\sim$1.0 $\mu$m with decreasing wavelength in a manner reminiscent of Rayleigh scattering, 
However, due to the large cross sections implied, the culprit can only be a haze or cloud.  
Note that these transit data can't distinguish between absorption and scattering, though scattering
is the more likely for most plausible haze materials and particle sizes. Scattering
is also indicated by the near lack of evidence for absorbing particulates in its secondary eclipse 
emission spectrum \cite{grillmair}. Together, these data suggest that a scattering haze layer 
at altitude is obscuring the otherwise distinctive spectral features of the spectroscopically 
active atmospheric constituents.

Transit spectra for the mini-Neptune GJ 1214b have been taken by many groups, but the results 
until recently have been quite ambiguous concerning possible distinguishing spectral features \cite{howe_gj1214b}.
In principle, there are diagnostic water features at $\sim$1.15 and 1.4 $\mu$m.  
However, Kreidberg et al. \cite{kreidberg} using the WFC3 on the Hubble Space Telescope,
have demonstrated that from $\sim$1.1 to 1.6 $\mu$m its transit spectrum is 
$\sim$5$-$10 times flatter than a water-dominated atmosphere or the canonical molecular-hydrogen(H$_2$)-dominated atmosphere 
with a solar abundance of water (oxygen), respectively (Figure 2).  Flatness could indicate that
the atmosphere has no scale height (eq. \ref{sigma_r}) (due, for example to a high mean molecular weight, $\mu$), 
or herald the presence yet again of a thick haze layer obscuring the molecular features. 
Not surprisingly, a pan-chromatically obscuring haze layer is currently the front runner. 

Lest one think that hazes completely mask the molecules of exoplanet atmospheres,
Deming et al. \cite{deming_haze} have published transit spectra of HD 209458b (Figure 3) and XO-1b 
that clearly show the water feature at $\sim$1.4 $\mu$m.  However, the expected 
accompanying water feature at $\sim$1.15 $\mu$m is absent.  The best interpretation is that this feature 
is suppressed by the presence of a haze with a continuum, though wavelength-dependent, interaction
cross section that trails off at longer wavelengths. The weaker apparent degree of 
suppression in these exoplanet atmospheres might suggest that their hazes are 
thinner or deeper (at higher pressures) than in HD 189733b.  Physical models explaining
this behavior are lacking.

So, the only atmospheric species that have clearly been identified in transit are H$_2$O, 
Na, K, and a ``haze".  Molecular hydrogen is the only gas with a low enough $\mu$ to
provide a scale height great enough to explain the detection in transit of 
any molecular features (eq. \ref{sigma_r}) in a hot, irradiated atmosphere, and I would
include it as indirectly indicated.  However, carbon monoxide (CO), carbon dioxide (CO$_2$),
ammonia (NH$_3$), nitrogen gas (N$_2$), acetylene (C$_2$H$_2$), ethylene (C$_2$H$_4$), 
phosphine (PH$_3$), hydrogen sulfide (H$_2$S), oxygen (O$_2$), ozone (O$_3$), nitrous oxide (N$_2$O), 
and hydrogen cyanide (HCN) have all been proferred as exoplanet atmosphere gases.  Clearly, 
the field is in its spectroscopic infancy. Facilities such as next-generation 
ground-based telescopes (Extremely-Large Telescopes, ELTs) and space-based telescopes such as the 
James Webb Space Telescope (JWST) \cite{deming_JWST}, or a dedicated exoplanet space-based 
spectrometer, will be vital if transit spectroscopy is to realize its true potential 
for exoplanet atmospheric characterization.  JWST in particular will have
spectroscopic capability from $\sim$0.6 to $\sim$25 $\mu$m and will be sensitive to
most of the useful atmospheric features expected in giant, neptune, and sub-neptune exoplanets.
It may also be able to detect and characterize a close-in earth or super-earth around a small nearby 
M star.
 
There are a number of theoretical challenges that must be met before transit
data can be converted into reliable knowledge.  Such spectra probe the terminator region 
of the planet that separates the day and night sides. They sample the transitional
region between the hotter day and cooler night of the planet where the 
compositions may be changing and condensates may be forming.  Hence, the compositions 
extracted may not be representative even of the bulk atmosphere.  Ideally, one would
want to construct dynamical 3D atmospheric circulation models that couple
non-equilibrium chemistry and detailed molecular opacity databases with 
multi-angle 3D radiation transfer. Given the emergence of hazes and clouds 
as potentially important features of exoplanet atmospheres, a meteorologically 
credible condensate model is also desired.  We are far from the latter \cite{marley_2013}, and the former
capabilities are only now being constructed, with limited success \cite{showman}. 
The dependence of transit spectra on species abundance is weak, making it difficult
now to derive mixing ratios from transit spectra to better than a factor of ten to one 
hundred.  Though the magnitude of the variation of apparent radius with wavelength depends upon
atmospheric scale height, and, hence, temperature, the temperature$-$pressure profile
and the variation of abundance with altitude are not easily constrained.  To obtain
even zeroth-order information, one frequently creates isothermal atmospheres with
chemical equilibrium or uniform composition.  Current haze models are ad hoc, and 
adjusted a posteriori to fit the all-too-sparse and at times ambiguous data.  
To justify doing better will require much better, and higher-resolution measured 
spectra \cite{burrows_pnas}.  

Data at secondary eclipse require a similar modeling effort, but probe the integrated
flux of the entire dayside.  Hence, a model that correctly incorporates the effects
of stellar irradiation (``instellation") and limb effects is necessary.  Moreover, 
the flux from the cooling planetary core, its longitudinal/latitudinal variation, 
and a circulation model that redistributes energy and composition are needed. 
Most models employed to date use a representative 1D (planar) approximation
and radiative and chemical equilibrium for what is a hemispherical region that
might be out of chemical equilibrium (and slightly out of radiative equilibrium).  
The emission spectra of the dayside depend more on the absorptive opacities, 
whereas transit spectra depend on both scattering and absorption opacities.  
Hence, if the haze inferred in some transit spectra is due predominantly 
to scattering, its effect on secondary eclipse spectra will be minimal, making
it a bit more difficult to use insights gained from one to inform the modeling 
of the other.

Many giant exoplanets, and a few sub-Neptunes, have been observed at secondary eclipse,
but the vast bulk of these data are comprised of a few photometric points per planet.
The lion's share have been garnered using {\it Spitzer}, the HST, or large-aperture
ground-based telescopes, and pioneering attempts to inaugurate this science
were carried out by Deming et al. \cite{hd209_deming_24} and Charbonneau et 
al. \cite{detection_tres1}.   Photometry, particularly if derived using techniques subject 
to systematic errors, is ill-suited to delivering solid information on composition, thermal
profiles, or atmospheric dynamics.  The most one can do with photometry at secondary eclipse
is to determine rough average emission temperatures, and perhaps reflection albedos in the 
optical.  Temperatures for close-in giant exoplanet atmospheres from $\sim$1000 to $\sim$3000 K
have in this way been determined.  Of course, the mere detection of an exoplanet is a victory, and the efforts
that have gone into winning these data should not be discounted.  Nevertheless, 
with nearly fifty such campaigns and detections ``in the can", one has learned that it 
is only with next-generation spectra using improved (perhaps dedicated) spectroscopic
capabilities that the desired thermal and compositional information will be forthcoming.

One of the few reliable compositional determinations at secondary eclipse obtained 
so far is for the dayside atmosphere of HD 189733b using the now-defunct IRS 
spectrometer onboard {\it Spitzer} \cite{grillmair}.  This very-low-resolution 
spectrum nevertheless provided a $\sim$3-$\sigma$ detection of water at $\sim$6.2 $\mu$m.  
There are other claims in the literature to have detected molecules
at secondary eclipse, but many are less compelling, and previous claims to have 
detected water using photometry alone at secondary eclipse are very 
model-dependent \cite{hd209_tres1_burrows_2005}. It is only with
well-calibrated spectra that one can determine with confidence the presence 
in any exoplanet atmosphere of any molecule or atom.

\section{Winds from Planets}
\label{wind}

The existence of what are now somewhat contradictorily called ``hot Jupiters"
has since the discovery of 51 Peg b in 1995 been somewhat of a puzzle.  They likely 
cannot form so close to their parent star and must migrate in by some process from 
beyond the so-called ice line.  In such cold regions, ices can form and accumulate 
to nucleate gas giant planet formation. Subsequent inward migration could be driven early 
in the planet's life by gravitational torquing by the protostellar/protoplanetary 
disk or by planet-planet scattering, followed by tidal dissipation in the planet 
(which circularizes its orbit).  However, once parked at between $\sim$0.01 and 0.1 A.U.
from the star, how does the gaseous planet, or a gaseous atmosphere of a smaller planet, 
survive evaporation by the star's intense irradiation during perhaps billions of years 
seemingly in extremis? The answer is that for sub-Neptunes and rocky planets their 
atmospheres or gaseous envelopes may indeed not survive, but for more massive gas giants the 
gravitational well at their surfaces may be sufficiently deep.  Nevertheless,
since the first discoveries evaporation has been an issue \cite{bur_wind}.  The atmospheres of Earth and 
Jupiter are known to be evaporating, though at a very low rate.  But what of a hot
Jupiter under $\sim$10$^4\times$ the instellation experienced by Jupiter?

The answer came with the detection by Vidal-Madjar et al. \cite{vidal} of a wind
from HD 209458b. Using the transit method, but in the ultra-violet around the Lyman-$\alpha$
line of {\it atomic} hydrogen at $\sim$0.12 $\mu$m, these authors measured a transit depth of $\sim$15 \%!
Such a large depth implies a planet radius greater than 4 R$_{\rm J}$, which is not
only far greater than what was inferred in the optical, but beyond the tidal Roche radius.
Matter at such a distance is not bound to the planet and the only plausible explanation
is that a wind was being blown off the planet.  The absorption cross sections in the ultraviolet are huge,
so the matter densities necessary to generate a transverse/chord optical depth of one 
are very low, too low to affect the optical and infrared measurements.  The upshot
is the presence of a quasi-steady planetary wind with a mass loss rate of 10$^{10-11}$ gm s$^{-1}$.
At that rate, HD 209458b will lose no more than $\sim$10\% of its mass in a Hubble time.

Since this initial discovery, winds from the hot Jupiters HD 189733b \cite{vidal_189} 
and WASP-12b \cite{wasp_wind} and from the hot Neptune GJ 436b \cite{kulow} 
have been discovered by the UV transit method and partially characterized. In all cases,
the tell-tale indicator was in atomic hydrogen. Mass loss rates have been estimated \cite{ehren_wind},
and in the case of WASP-12b might be sufficient to completely evaporate the giant within 
as little as $\sim$1 gigayear.  The presence of atomic hydrogen 
implies the photolytic or thermal breakup of the molecular hydrogen, so these data 
simultaneously suggest the presence of both H and H$_2$. Linsky et al. \cite{linsky1} detected
ionized carbon and silicon in HD 209458b's wind and Fossati et al. \cite{wasp_wind}
detected ionized magnesium WASP-12b's wind, but the interpretation of the various ionized species
detected in these transit campaigns is ongoing.

The theoretical challenges posed by planetary winds revolve in part around the driver.  Is it energy-limited
UV and X-ray flux from the parent star, or heating by the integral intercepted stellar light?
In addition, in the rotating system of the orbiting planet, what ingress/egress asymmetries
in the morphology of the wind are there?  There are indications that Coriolis forces on planet 
winds are indeed shifting the times of ingress and egress.  What is the effect of planet-star wind interactions? 
There are suggestions of Doppler shifts of lines of the UV transit data that arise from
planet wind speeds, but how can we be sure? How is the material for the wind replenished from 
the planet atmosphere and interior?  And finally, what is the correspondence between
the UV photolytic chemistry in the upper reaches of the atmosphere that modifies 
its composition there and wind dynamics? This is a rich subject tied to many sub-fields of science,
and is one of the important topics to emerge from transit spectroscopy.

\section{Phase Light Curves and Planet Maps}
\label{light_map}

As a planet traverses its orbit, its brightness as measured at the Earth at a given wavelength
varies with orbital phase.  A phase light curve is comprised of 1) a reflected component
that is a stiff function of star-planet-Earth angle and is most prominent in the 
optical and UV and 2) a thermal component that more directly depends upon the 
temperature and composition of the planet's atmosphere and their longitudinal variation
around the planet and is most prominent in the near- and mid-infrared.  Hence, a phase light curve is 
sensitive to the day-night contrast and is a useful probe of planetary atmospheres 
\cite{marley_1999,sudarsky_albedo,burrows_2004,barman_2005,analytic_albedos}. Note that
the planet/star contrast ratio is largest for large exoplanets in the closest orbits, 
so hot Jupiters currently provide the best targets.

In the optical, there has been some work to derive the albedo \cite{marley_1999,sudarsky_albedo}, or reflectivity, 
of close-in exoplanets, which is largest when there are reflecting clouds and smallest when the atmosphere is absorbing.
In the latter case, thermal emission at high atmospheric temperatures can be mistaken for reflection,
so detailed modeling is required.  In any case, {\it Kepler}, with its superb photomteric sensitivity,
has been used to determine optical phase curves \cite{esteves} of a few exogiants in the 
{\it Kepler} field and the MOST microsatellite has put a low upper limit on the optical albedo 
of HD 209458b \cite{hd209_albedo_rowe,hd209_albedo_burrows}, but much remains to be done 
to extract diagnostic optical phase curves and albedos for exoplanets.

Interesting progress has been made, however, in the thermal infrared.  Using {\it Spitzer} at 8 $\mu$m,
Knutson et al. \cite{knutson_189_map} not only derived a phase light curve for HD 189733b, but derived
a crude thermal map of its surface.  By assuming that the thermal emission pattern over the planet 
surface was fixed during the observations, they derived the day-night brightness contrast (translated into
a brightness temperature at 8 $\mu$m) and a longitudinal brightness temperature distribution.  In particular,
they measure the position of the ``hot spot."  If the planet is in synchronous rotation 
(spin period is the same as the orbital period), and there are not equatorial winds to advect heat 
around the planet, one would expect the hot spot to be at the substellar point.  The light curve
would phase up with the orbit and the peak brightness would occur at the center of secondary eclipse.
However, what they observed was a shift ``downwind" to the east by $\sim$16$^{\circ}$$\pm$6$^{\circ}$.
The most straightforward interpretation is that the stellar heat absorbed by the planet is advected 
downstream by superrotational flows such as are observed on Jupiter itself
before being reradiated.  Moreover, these data indicate that since the measured day-night brightness 
temperature contrast was only $\sim$240 K the zonal wind flows driven by stellar irradiation indeed 
carry heat to the night side, where it is radiated at a detectable level. Hence, these data point 
to atmospheric dynamics on the exoplanet HD 189733b qualitatively (though not quantitatively) 
in line with theoretical expectations \cite{showman}.

For HD 189733b, this work has been followed up using {\it Spitzer} at 3.6 and 4.5 $\mu$m \cite{hd189_2012_new} and 
in a competing effort a more refined map has been produced \cite{majeau}.  Infrared phase curves for the giants
HD 149026b \cite{hd149026b}, HAT-P-2b \cite{hatp2_lewis}, and WASP-12b \cite{wasp12_cowan}, among other exoplanets, 
have been obtained.  However, one of the most intriguing phase curves was obtained by Crossfield et al. 
\cite{crossfield_ups} using {\it Spitzer} at 24 $\mu$m for the {\it non}-transiting planet $\upsilon$ And b (Figure 4).  These
authors found a huge phase offset of $\sim$80$^{\circ}$, for which a cogent explanation is still lacking. 
The closeness of this planet to Earth could compensate in part for the fact that it is not transiting to
allow sufficient photometric accuracy without eclipse calibration to yield one of the few non-transiting light curves.
All these efforts collectively demonstrate the multiple, at times unanticipated and creative, methods being
employed by observers seeking to squeeze whatever information they can from exoplanets.

Theoretical models for light curves have been sophisticated, but theory and measurement 
have not yet meshed well.  Both need to be improved.  Models need to 1) improve their 
treatment of hazes and clouds that could reside in exoplanet atmospheres and will
boost reflection albedos significantly; 2) incorporate polarization to 
realize its diagnostic potential \cite{analytic_albedos,Seager_polar};
3) constrain the possible range of phase functions to aid in retrievals; 4) embed the effects 
of variations in planet latitude and longitude in the analysis protocols; 5) provide observational diagnostics 
with which to probe atmospheric pressure depths, particularly using multi-frequency data; 6) be 
constructed as a function of orbital eccentricity, semi-major axis, and inclination; and      
7) span the wide range of masses and compositions the heterogeneous class of exoplanets in likely 
to occupy.  Accurate spectral data with good time coverage from the optical to the 
mid-infrared could be game-changing, but theory needs to be ready with useful 
physical diagnostics.

\section{High Spectral Resolution Techniques}
\label{high_res}

The intrinsic dimness of planets under the glare of stars renders high-resolution, 
pan-chromatic spectral measurements difficult, if desirable.  However, ultra-high
spectral resolution measurements using large-aperture ground-based telescopes,
but over a very narrow spectral range and targeting molecular band features 
in a planet's atmosphere otherwise jumbled together at lower resolutions, 
has recently been demonstrated.  Snellen et al. \cite{hd209_snellen_winds} 
have detected the Doppler variation due to HD 209458b's orbital motion 
of carbon monoxide features near $\sim$2.3 $\mu$m.  The required 
spectral resolution ($\frac{\lambda}{\Delta\lambda}$) was $\sim$10$^5$ and 
the planet's projected radial velocity just before and just after primary 
transit changed from +15 km s$^{-1}$ to -15 km s$^{-1}$. This is consistent 
with the expected circular orbital speed of $\sim$140 km s$^{-1}$ and provides 
an unambiguous detection of CO.  What is more, this team was almost able to 
measure the zonal wind speeds of air around the planet, estimated theoretically
to be near $\sim$1 km s$^{-1}$, thereby demonstrating the potential of such a novel
technique to extract weather features on giant exoplanets.  The same basic method
has been applied near primary transit to detect CO \cite{dekok} and H$_2$O 
\cite{birkby} in HD 189733b. Carbon monoxide is detected in Jupiter and was 
thermochemically predicted to exist in abundance in the atmospheres of hot 
Jupiters \cite{burrows_sharp}, but its actual detection by this method is 
impressive.

In fact, the same technique has been succesfully applied in the CO band 
to the non-transiting planet $\tau$ Boo b \cite{brogi} and for the wide-separation 
giant planet/brown dwarf $\beta$ Pictoris b \cite{snellen_betapic}, verifying 
the presence of CO in both their atmospheres.  Finally, using a related technique 
Crossfield et al. \cite{crossfield_luhman} have been able to conduct 
high-resolution ``Doppler imaging" of the closest brown dwarf known (Luhman 16B).
By assuming that the brown dwarf's surface features are frozen during the observations
and that it is in solid-body rotation, tiling its surface in latitude and longitude 
they were able to back out surface brightness variations from the variations of its flux 
and Doppler-shift time series. By this means, they have mapped surface spotting that 
may reflect broken cloud structures (Figure 5). 

In support of such measurments, theory needs to refine its modeling of planet surfaces,
zonal flows and weather features, three-dimensional heat redistribution and velocity 
fields, and temporal variability. Currently, most 3D general circulation models 
do not properly treat high Mach number flows, yet they predict zonal 
wind Mach numbers of order unity.  There are suggestions that magnetic fields
affect the wind dynamics and heating in the atmosphere, but self-consistent multi-dimensional
radiation magnetohydrodynamic models have not yet been constructed.

This series of measurements of giant exoplanets and brown dwarfs 
using high-resolution spectroscopy focused on narrow molecular features 
emphasizes two important aspects of exoplanet research. The first is that 
observers can be clever and develop methods unanticipated in Roadmap documents 
and Decadal Surveys.  The second is that with the next-generation of ground-based 
ELTs equiped with impressive spectrometers astronomers may 
be able to measure and map some exoplanets without employing the high-contrast imaging
techniques that are now emerging to compete and to which I now turn.

\section{High-Contrast Imaging}
\label{imaging}

Before the successful emergence of the RV and transit methods, astronomers
expected high-contrast direct imaging that separated out the light of planet and the star,
and provided photometric and spectroscopic data for each, would be the 
leading means of exoplanet discovery and characterization.  A few wide-separation
brown dwarfs and/or super-Jupiter planets were detected by this means, but the
yield was meager. The fundamental problem is two-fold: 1) the planets are intrinsically dim,
and 2) it is difficult to separate out the light of the planet from under the glare of the star
for planet-star separations like those of the solar system.  Imaging systems need to suppress the 
stellar light scattered in the optics that would otherwise swamp the planet's signature.  The planet/star contrast
ratio for Jupiter is $\sim$10$^{-9}$ in the optical and $\sim$10$^{-7}$ in the mid-infrared.
For Earth, the corresponding numbers are $\sim$10$^{-10}$ and $\sim$10$^{-9}$.  These numbers 
are age, mass, orbital distance, and star dependent, but demonstrate the challenge. What is more,
contrast capabilities are functions of planet-star angular separation, restricting
the orbital space accessible.    

However, high-contrast imaging is finally emerging to complement other methods.  It
is most sensitive to wider-separation ($\sim$10$-$200 AU), younger, giant exoplanets (and brown dwarfs),
but technologies are coming online with which to detect older and less massive exoplanets 
down to $\sim$1 AU separations for nearby stars ($\le$10 parsecs) \cite{GPI,SPHERE,spergel,suzuki,burrows_2005}.
Super-Neptunes around M dwarfs might soon be within reach.  Using direct imaging, Marois et al. 
\cite{marois,marois_e} have detected four giant planets orbiting the A star HR 8799 (HR 8799b,c,d,e) 
and Lagrange et al. \cite{betapic} have detected a planet around the A star $\beta$ Pictoris.  
The contrast ratios in the near infrared is $\sim$10$^{-4}$, but capabilities near 10$^{-5}$
have been achieved and performance near 10$^{-7}$ is soon anticipated \cite{GPI,SPHERE}.
One of the results to emerge from the measurements of both the HR 8799 and $\beta$-Pic planets
is that to fit their photometry in the near-infrared from $\sim$1.0 to $\sim$3.0 $\mu$m thick clouds,
even thicker than seen in L dwarf brown dwarf atmospheres, are necessary \cite{hr8799_madhu}.
This (re)emphasizes the theme that the study of hazes and clouds (nephelometry) has emerged 
as a core topic in exoplanet studies. 

One of the most exciting recent measurements via direct imaging was by Konopacky 
et al. \cite{konopacky} of HR 8799c. Using the OSIRIS spectrometer on the 10-meter Keck II 
telescope, they obtained unambiguous detections between $\sim$1.95 and $\sim$2.4 $\mu$m 
of both water and carbon monoxide in its $\sim$1000 K atmosphere (Figure 6). This 
$\frac{\lambda}{\Delta\lambda} = 4000$ spectrum is one of the best 
obtained so far, but was enabled by the youth ($\sim$30 million years), wide-angular 
separation, and large mass ($\sim$5$-$10 M$_{\rm J}$) of the planet.  

Improvement in theory needed to support direct imaging campaigns mirror those
needed for light curves, but are augmented to include planet evolution modeling 
to account for age, metallicity/composition, and mass variations.  Most high-contrast
instruments are focused on the near-infrared, so cloud physics and near-infrared
line lists for likely atmospheric constituents will require further work. The
reader will note that the vast majority of observations and measurements of
exoplanet atmospheres has been done for giants.  There are a few for sub-Neptunes 
and super-Earths, but high-contrast measurments of earths around G stars like the Sun 
is not likely in the near future \cite{kalten_2007,ehrenreich}.  The planet/star contrast ratios are just too low, 
though earths around M stars might be within reach if we get lucky. For now, giants and Neptunes
are the focus, as astronomers hone their skills for an even more challenging future.

\section{What We Know about Atmospheric Compositions}
\label{comp}

To summarize, the species we have, without ambiguity, discovered to date in exoplanet atmospheres are:
H$_2$O, CO, Na, K, and H (H$_2$), with various ionized metals indicated 
in exoplanet winds.  Expected species, but as yet undetected, include: NH$_3$, 
CH$_4$, N$_2$, CO$_2$, H$_2$S, PH$_3$, HCN, C$_2$H$_2$, C$_2$H$_4$, O$_2$, O$_3$, and N$_2$O.
The nature of the hazes and clouds inferred is at yet unknown.  The atmospheres probed have temperatures from
$\sim$600 K to $\sim$3000 K.  Good spectra are the essential requirements for unambiguous 
detection and identification of molecules in exoplanet atmospheres, and these have been rare.
Determining abundances is also difficult, since to do so requires not only good spectra, but 
reliable models.  Errors in abundance retrievals of more than an order of magnitude are likely,
and this fact has limited the discussion of abundances in this paper.  

Nevertheless, with the construction of ground-based ELTs, the various campaigns of direct 
imaging \cite{GPI,SPHERE,suzuki}, the launch of JWST, the possible
launch of WFIRST-2.4/AFTA \cite{spergel}, and the various ongoing campaigns 
with HST and {\it Spitzer} and extant ground-based facilities, the near-term 
future of exoplanet atmospheric characterization promises to be even more exciting than its past.

\begin{acknowledgments}
The author acknowledges support in part under
Hubble Space Telescope grants HST-GO-12181.04-A, HST-GO-12314.03-A, HST-GO-12473.06-A, and HST-GO-12550.02,
and Jet Propulsion Laboratory/Spitzer Agreements 1417122, 1348668, 1371432, 1377197, and 1439064.
\end{acknowledgments}

\newcounter{firstbib}

\renewcommand{\refname}{Reference 1}

\newpage

% figure 1
\begin{figure}
\includegraphics[height=0.38\textheight,angle=0]{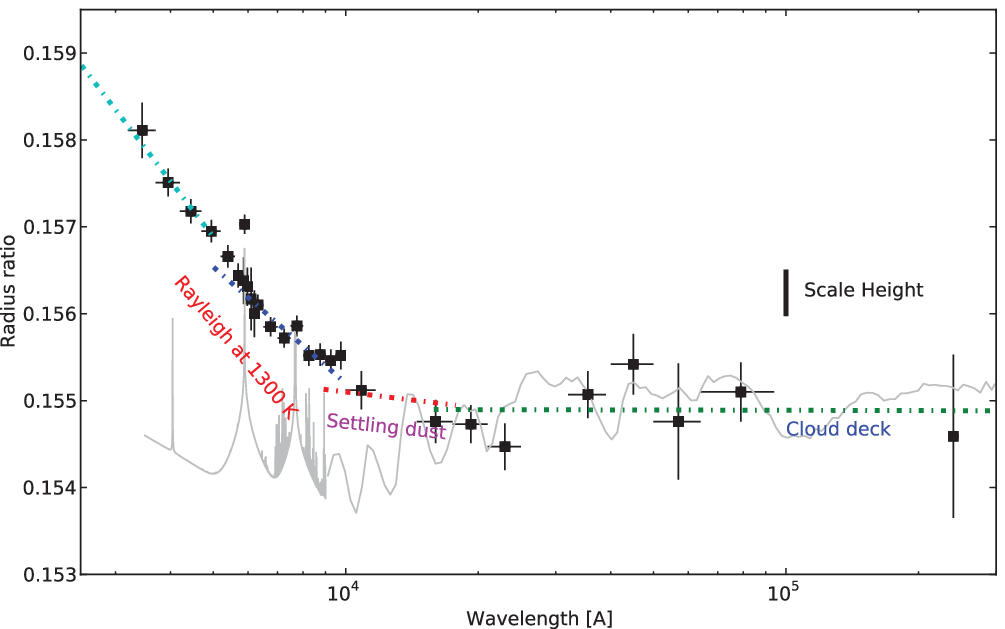}
\caption{The planet/star radius ratio versus wavelength in \AA\ for the giant exoplanet HD 189733b is here depicted in transit.  
The black dots are the data points and the lines are models.  The grey line is an example 
spectrum without a haze.  Reprinted with permission from reference \cite{pont2}.
\label{fig1}}
\end{figure}

\clearpage 

% figure 2
\begin{figure}
\includegraphics[height=0.38\textheight,angle=0]{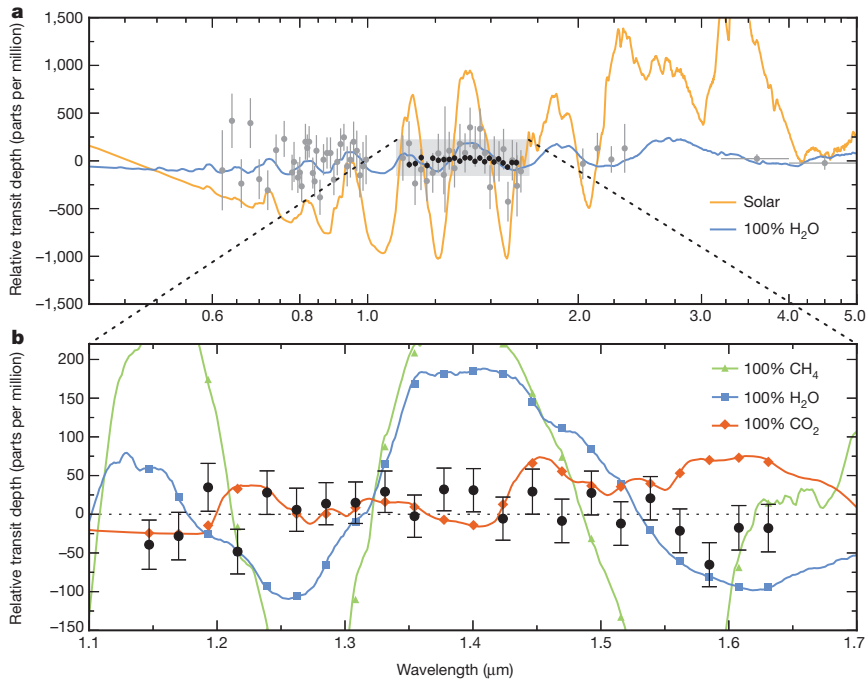}
\caption{This figure portrays the relative depth of the transit of the 
sub-Neptune GJ 1214b versus wavelength from 1.1 to 1.7 $\mu$m.  
The colored lines on both top and bottom panels are various transit spectral models 
without a haze.  The data (black dots) are effectively flat, ruling out all
models shown and suggesting a veiling haze. Reprinted with permission from reference \cite{kreidberg}.
\label{fig2}}
\end{figure}

\clearpage

% figure 3
\begin{figure}
\includegraphics[height=0.38\textheight,angle=0]{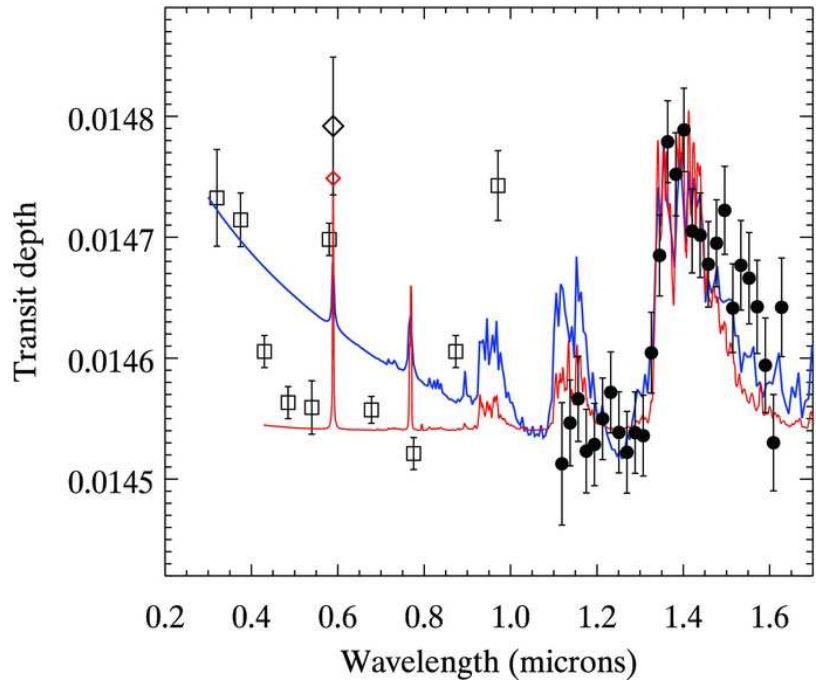}
\caption{The transit depth spectrum (black points from reference \cite{deming_haze}) 
versus wavelength from 0.2 to 1.7 $\mu$m of the hot Jupiter HD 209458b is shown here.
The presence of water is demonstrated by the feature at 1.4-$\mu$m, but the 
corresponding $\sim$1.15 $\mu$m feature is absent. The best explanation is that 
the latter is supporessed by haze scattering.  Not obvious here is the
fact that even the 1.4-$\mu$m feature is muted with respect to non-haze models.  
The two colored curves are representative model spectra with different levels of haze. 
Reprinted with permission from reference \cite{deming_haze}.
\label{fig3}}                                                                                                                                                \end{figure}

\clearpage 

% figure 4
\begin{figure}
\includegraphics[height=0.25\textheight,angle=0]{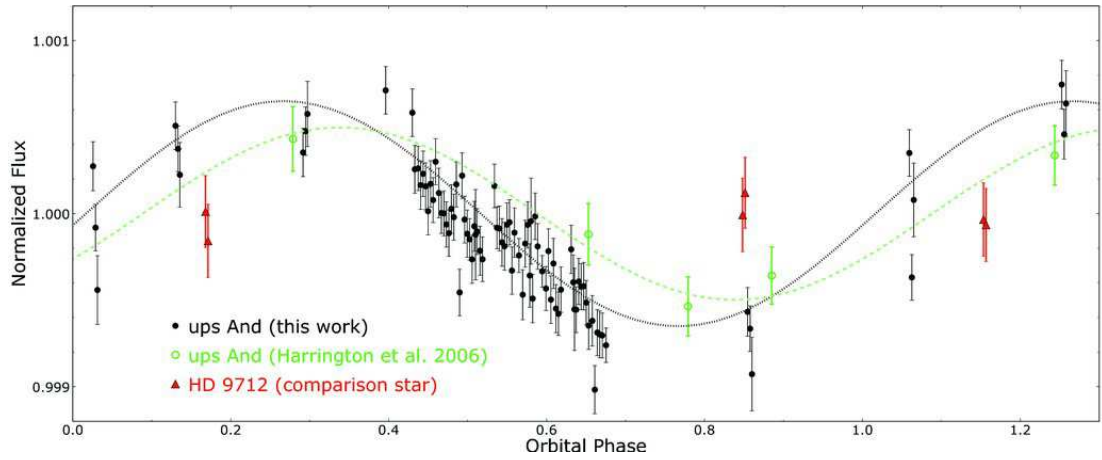}
\caption{Here, the measured light curve at a wavelength of 24 $\mu$m (black dots with 
error bars) of $\upsilon$ And b is depicted versus orbital phase.  The thin 
black dotted curve is the authors' best fit, showing a phase offset of $\sim$80$^{\circ}$ (22\% of a circuit).
Reprinted with permission from reference \cite{crossfield_ups}.
\label{fig4}}                                                                                                                                                \end{figure}

\clearpage 

% figure 5
\begin{figure}
\includegraphics[height=0.38\textheight,angle=0]{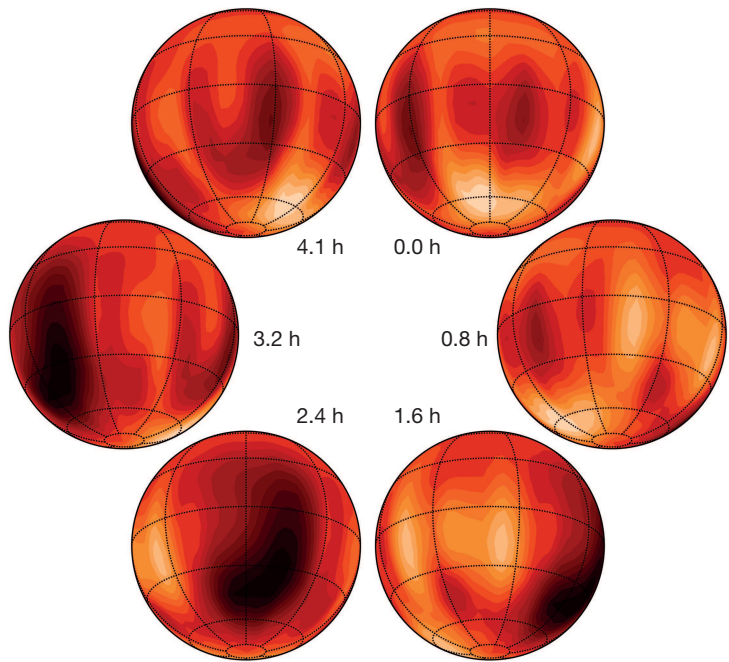}
\caption{These surface maps are obtained by Doppler Imaging and depict different 
epochs during the rotation of the brown dwarf Luhman 16B. Large-scale cloud
inhomogeneities are suggested. The rotation period of the brown dwarf is 4.9 hours.
Reprinted with permission from reference \cite{crossfield_luhman}.
\label{fig5}}
\end{figure}

\end{article}

\end{document}